\documentclass[a4paper,12pt]{article}
\usepackage{color}
\usepackage[utf8x]{inputenc}
\usepackage{graphicx}
\usepackage{wrapfig}
\usepackage{footnote}
\usepackage{caption}
\def\clr{\color{black}}
\usepackage[affil-it]{authblk}

\title{Communities of dense weighted networks: MicroRNA co-target network as an example}
\author{Mahashweta Basu \thanks{Electronic address: \texttt{mahashweta.basu@saha.ac.in}}} 

\date{}
\providecommand{\keywords}[1]{\textbf{\textit{Keywords :}} \textit{#1}}
\affil{Condensed Matter Physics, Saha Institute of Nuclear Physics, 1/AF Bidhannagar, Kolkata 700064, India.}

\begin{document}
\maketitle
\begin{abstract}
Complex networks are intrinsically modular. Resolving small modules is particularly difficult when the
network is densely connected; wide variation of link weights invites additional complexities.
In this article we present an algorithm to detect community structure in densely connected 
weighted networks. First,  modularity of the network is calculated by  
erasing the links having weights smaller than a cutoff $q.$  
Then one takes all the disjoint components obtained at $q=q_c,$ where the modularity is maximum, and
modularize the components individually using Newman Girvan's algorithm for weighted networks. We show, taking microRNA (miRNA) co-target network of \textit{Homo sapiens} as an example, that this algorithm could reveal miRNA modules which are known to be relevant in biological context.
\end{abstract}

\maketitle

\section{Introduction}
Networks, a set of nodes or vertices joined pairwise by links or edges, are commonly used for
describing sociological (scientific collaborations \cite{collab} 
and acquaintance networks \cite{friendship}), biological (proteins interactions, 
genes regulatory, food webs, neural networks, metabolic networks), technological 
(Internet and the web) and communication (airport \cite{airport}, 
road \cite{road1,road2}, and railway network \cite{Indian_railway,china_railway}) systems.
The topological properties of these complex networks \cite{barabasirev,barabasibook} 
help in identifying underlying community structures \cite{NG_betw1}, network 
motifs \cite{urialon}, connectivity \cite{israelattack,packagetransfer} and several other properties 
\cite{otherbook}. The links of a network can also be weighted.
Some of the networks are associated with links of varying strengths \cite{weightednet,weightednet1,weightednet2}
represented by link weights. The topological properties of weighted networks \cite{weightednet3,weightednet4} are 
quite different and their study requires additional care. In particular when link weights 
vary in a wide range, one need to identify suitably the irrelevant links and ignore them 
to simplify  the network \cite{susmita}.

Most networks in nature, whether weighted or not, exhibit community (or modular) structures.
Detection of communities in the complex networks provide invaluable information
on the underlying synergism. Nodes which belong to a particular module
are more than likely to function together for some common cause; being able to unravel
such communities help in identifying functional properties of the network. For example 
in social networks \cite{social_wasserman}, communities observed  are based on 
interests, age, profession of the people. Similarly, communities reflects the themes 
of the web-pages in World Wide Web, related papers on a single topic
in citation networks  \cite{citation}, subsystems within ecosystems \cite{foodweb1,foodweb2} 
in food webs, and  it may relate to functional groups \cite{holme,ravasz} in cellular and 
metabolic networks. 

To identify the modular structures of complex networks, 
several algorithms \cite{kernighan,fiedler,pothen,scott} are developed recently.
But most of these methods are context based and a unique algorithm which could work 
universally is still out of reach. Recently Newman and Girvan has proposed a couple of methods 
\cite{NG_betw1,NG_betw2,N2004} to detect the modules but they take high computational 
time for large networks. Later, a faster algorithm \cite{A2004} is being put 
forward by the same authors, based on maximization of modularity $M$ defined as the number 
of edges present within the groups minus the expected number in an equivalent random network.
According to this algorithm, best partition of a network is the one which has maximum modularity $M.$ 
This modularization method \cite{A2004} is further generalized to include weighted networks \cite{N_wt}.

Newman Girvan's modularization algorithms (NGM), though widely used for finding modules of 
both weighted and 
unweighted networks, has some shortcomings \cite{fortunato}.  It was argued that 
modularity   maximization   algorithm  can resolve the network upto a scale  that 
depends on the  total number  of links $l;$ a module having  more than $\sqrt{l/2}$ links 
can  not be resolved {\clr even when   it is a 
clique and connected   to  external  modules   through just one link. }
Moreover the situation gets worse when substantial  number of 
small communities coexist with large ones.
This observation is also true for weighted networks \cite{MNP}.
Therefore modularity maximization uncovers  only large modules 
missing important substructures which are small.  
In this context, an clustering method has been proposed   recently  by 
Mookherjee \textit{et. al.} \cite{susmita}  in context of   microRNA   
co-target network  of human  which is  densely connected   by weighted links. 
The authors claimed  to   obtain  microRNA  clusters  which reveal  
biologically significant processes and pathways.  
This algorithm  also suffers  from   certain  short comings. First,  the 
method has  in-built  arbitrariness    in determining the  total  number 
of   clusters  and   then its 
sub-structures connected  by large-weighted links, if any,  remains 
undetectable. Details of  the algorithm and  its  shortcomings    are discussed  
in the next section. 

In this article we propose a new algorithm   in an  effort to  overcome 
these  shortcomings and to efficiently determine the communities of any dense 
weighted network. We demonstrate the  algorithm   using  the  microRNA co-target 
network  of \textit{Homo sapiens}  and compare the  modules    with  those 
obtained  by  NGM algorithm for weighted networks \cite{N_wt} and the  clustering 
algorithm \cite{susmita}.

\section{Clustering algorithm}
In a   recent article \cite{susmita}, Mookherjee  {\it et. al.} have proposed an algorithm  to
find  clusters of   miRNA co-target network of \textit{Homo sapiens}. MicroRNAs are short 
non-coding RNAs  which    usually  suppress  gene expression  in  
post-transcriptional level \cite{farh}.   Taking 
the  predicted targets of  $711$ miRNAs   of  \textit{Homo sapiens} 
from  Microcosm Target database \cite{Microcosm}, the authors  constructed  the  co-target
network by   joining   miRNAs  pairwise   by   weighted links. The   link  
weight $w$  corresponds to the number of   common  targets  of the  concerned pair.  
The network thus constructed consists of $711$ miRNAs (nodes)
and $252405$ edges. Since the network is fully connected,  it is evident that 
clusters    containing  less than   half the number of nodes  can  not  be 
resolved by  standard  algorithms \cite{NG_betw2,N_wt}.  To obtain    the  clusters 
of this  densely packed network  Mookherjee \textit{et. al.}
in \cite{susmita} have  adopted   the following  strategy.

The link weights of this network vary in a wide range: minimum being  $1$ 
and maximum $1253.$ Thus most links are considered irrelevant in determining 
the clusters. In an attempt to  simplify the network, links with weights 
smaller than a pre-defined cutoff value $q$ are erased; the resulting network
breaks into small disjoint components. Denoting, $N(q)$ as the number components   
the authors find that $N(q)$ does not increase substantially until $q$ reaches a 
threshold value $q^*$ and then it breaks quickly into large number of components 
(Fig. 2C in \cite{susmita}). Thus the network is optimally connected at $q^*=103$ 
where ${d \over dq}N(q)$ is maximum. Among all the {\clr components} obtained at 
$q^*=103,$ the largest one ${\cal G}$ contains $479$ miRNAs. A large fraction of 
miRNAs present in ${\cal G}$ are found to down regulate expression of genes 
involved in several genetic diseases. To explore how miRNAs are organized in 
${\cal G},$ $q$ is increased further until the {\clr total} number of components 
does not change much. At $q = 160,$ the subgraph ${\cal G}$ has $70$ components 
(called miRNA clusters) and $149$ lone miRNAs. Note that if we  consider all 
$711$ miRNAs, instead of $479$ miRNAs belonging to ${\cal G},$ the total number 
of clusters would have been $94$ (see Table \ref{table:compare} for details).   

Further, the authors have analyzed these $70$ clusters and claimed that they are 
biologically relevant -either pathway or tissue or disease specific. Note that, 
even though the targets are predicted based on sequence similarities, the microRNA 
clusters reveals functionality quite well; only about $11$ clusters are found to 
contain miRNAs of identical seed sequence. Thus it is suggestive that a group of 
miRNA, instead of individual ones, are involved in carrying out necessary functions.

{\it Limitations :} Although the cluster finding algorithm discussed in \cite{susmita} 
partitions the miRNA co-target network into several components which  provide significant 
information about  the functions  of  miRNA   clusters,  it suffers  from  certain limitations. 
Firstly, there exists few clusters containing a   large number  (as large as $47$)   of 
miRNAs;   such  large clusters  produce   significant   noise  in    identifying  
pathways and functions    from enhancement analysis.
 \begin{figure}[h]
\vspace*{.6cm}
  \begin{center}
   \includegraphics[width=5cm,bb=24 74 456 146]{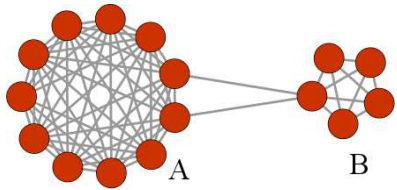}
  \end{center}
  \caption{The above network consists of two distinct modules (A and B) connected by only two links.}
\label{fig:limitation}
\end{figure} 
Secondly,     
if a  miRNA  cluster has two or more sub-structures which are connected by 
a few links having weights much larger than $q^*,$ it is beyond the scope of 
this algorithm to resolve them. For example the network in Fig. \ref{fig:limitation} 
clearly has two modules but {\clr weight of the} few links that joins the two modules 
are {\clr larger} than $q^*.$ Since the algorithm looks for disconnected components
of the graph, it is not possible to uncover these two obvious modules (A and B). Lastly
to reveal the sub-structures of a giant cluster ${\cal G},$ $q$ is increase to an 
arbitrary value (taken as $160$ in \cite{susmita}). In practice the actual number 
of clusters depends weakly on this choice, however it still  introduces an arbitrariness in the 
algorithm. All these shortcomings necessitates exploring other appropriate {\clr algorithms} 
for finding the community structure in dense weighted network.

\section{The proposed algorithm}
 In {\clr this section we  proposed an algorithm for finding modules of 
dense weighted networks.} The algorithm primarily consists of two steps -first, 
finding the major communities and second, extracting their sub-structures. 

\begin{figure}[h]
\centering
 \includegraphics[width=9cm,bb=14 14 976 501]{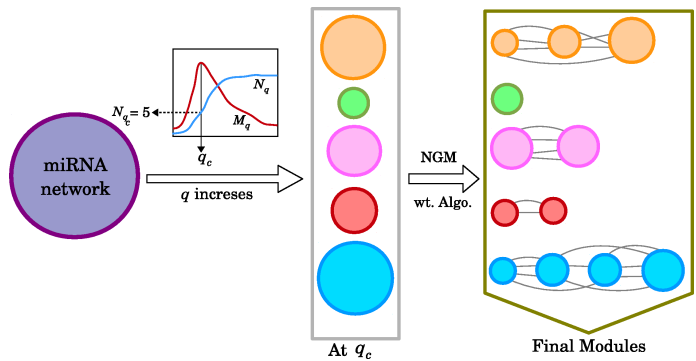}
 \caption{The outline of our algorithm that has been developed to find the
modules of the densely linked weighted network.}
 \label{fig:flowchart}
\end{figure}

\textit{Step I}: For finding the modular structures, we consider a weighted network 
which is densely connected. Let the network has $M$ nodes denoted by $i=1,2, \dots M$
and a connected pair of nodes $i$ and $j$ has non zero weight $W_{ij}.$  Thus, the 
network is represented by an adjacency matrix $W$ with elements 
{\clr \begin{equation}
W_{ij}= \left\{ 
\begin{array}{cc} 
w & {\rm if }~~ i  ~~ {\rm and}~~ j  ~~{\rm are ~~connected } \cr 0 &  {\rm otherwise} \end{array}\right..
\end{equation} }
We also assume that the network is densely connected. A preliminary simplification 
can be done following Ref. \cite{susmita}, where links with weights smaller than a 
pre-decided cutoff $q$ are erased. The resulting network {\clr thus breaks} up into
smaller disconnected components -say  $N(q)$  in  total. It is evident that $N(q)$ 
is the number of diagonal blocks of a  matrix $W^q$ with elements  
\begin{equation}
W_{ij}^q = \left\{ 
\begin{array}{cc} 
0 & \textrm{if~} W_{ij} < q  \cr W_{ij}&  \rm{ otherwise} \end{array}\right..
\nonumber
\end{equation}
Clearly  $ N(q)$  must strictly be  a non-decreasing function  with {\clr $N(q=0) = 1.$}

We proceed further to calculate the modularity of the concerned weighted network
for different values of $q.$ In general, if a network {\clr (weighted)} has $c$
partitions, one can calculate the modularity \cite{N_wt} from knowing the set of 
nodes which belong to each partition,  
\begin{equation}
 M={1\over 2m}\sum_{l=1}^c\sum_{ij}\left(W_{ij}- {k_i k_j\over 2m}  \right)S_{ij}^l 
\end{equation}
where $k_i= \sum_j W_{ij}$ represents sum of the
weights of the edges attached to node $i$ and $m=\sum_{i}k_{i}.$ The term $S_{ij}^l$ 
is $1$ only if vertices $i$ and $j$ belong to same group. For a given $q,$ we take 
the components as the modules (thus $c=N(q)$) and denote corresponding modularity 
as $M(q).$ Note that, unlike $N(q),$ the modularity $M(q)$ need not be a increasing 
function. A schematic plot of these functions are shown in Fig. \ref{fig:flowchart}. 
Since, large modularity is a feature of better community structure we choose the 
value $q_c$ where $M(q)$ takes the maximum value and then collect set of components 
obtained there for further analysis.

\textit{Step II} : The {\clr number of miRNAs present in each of the components,  
$i. e.,$ the component sizes} obtained at $q_c,$ are quite large. 
To get finner {\clr division} 
of these components we can increase the $q$ value further,  
then although we will get smaller sized groups but the value of $M(q)$  will decrease, which 
is not favorable. So keeping the value of $q$ fixed at $q_c$ where $M(q)$ is maximum, we find 
the further groups present in these individual components by using  NGM algorithm for 
weighted networks \cite{N_wt}. {\clr Taking the components one by 
one we then find their modules with help of NGM  algorithm for weighted network \cite{N_wt}, 
and accept the 
partition if the modularity value for this partition is positive or other wise we ignore it. }
Likewise we consider each of the components formed at $q_c$ for further partitioning. 
Collection of all the partitioned components of the network are then considered as the final modules
of the weighted network.

\section{Example case study}
 We demonstrate this algorithm    for   miRNA  co-target network of human, a   dense
 and weighted network   constructed    and studied by  Mookherjee 
 $et. al.$ \cite{susmita}.   MicroRNAs (miRNAs) are small 
single stranded $\sim22$nt long non-coding RNAs \cite{mirbook} that repress gene expression by 
binding $3'$-untranslated regions ($3'$ UTR) of  messenger RNA (mRNA) target transcripts, 
causing translational repression \cite{farh}. Being a secondary regulator, miRNAs usually repress
the gene expression marginally. Thus it is natural to expect that cooperative action of miRNAs 
are needed for alteration of any biological function or pathway. MicroRNA synergism 
has been a 
recent focus in biology for studying their regulatory effects in cell. Recent articles 
\cite{susmita,juan} have identified the assemblage of the miRNAs for performing various activities. In this view 
finding the small clusters or communities of the miRNAs that work together for regulatory 
functions is quite relevant.
For completeness,  first  we   describe the construction of miRNA co-target network briefly and then proceed  for  obtaining its modules  
using  the  algorithm discussed here.

\begin{figure}
\centering
 \includegraphics[width=13cm,bb=14 54 1020 495]{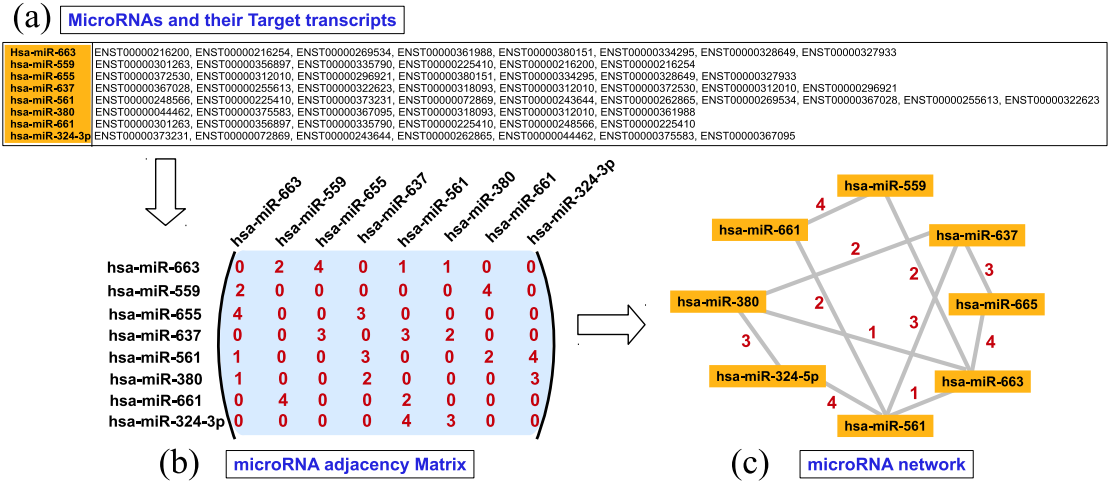}
 \caption{Construction of miRNA co-target network. (a) A representative data for $8$ miRNAs and  their targets transcripts. (b) The adjacency matrix ($W$) with elements $W_{ij}$ corresponding to
the number of common target transcripts. (c) The miRNA co-target network, where the miRNAs are represented as nodes which are connected by links having weight $W_{ij}.$}
 \label{fig:dataset}
\end{figure}

\subsection{Construction of miRNA co-target network}
The miRNAs which act as secondary regulators can target more than one mRNA transcripts
and a transcript can also be targeted by many miRNAs. Computationally predicted targets of miRNAs
for different species are available in Microcosm Target database \cite{Microcosm}.
For constructing the miRNA network the targets of miRNAs  are collected from the above mentioned database.
The data predicts $34788$ targets for $711$ miRNAs for \textit{Homo sapiens}.

The miRNA co-target network is constructed by considering miRNAs as nodes, and a link with weight $w$ 
is connected between two miRNAs if they both target $w$ number of same target transcripts. {\clr The 
detailed procedure for constructing the miRNA co-target network is shown in Fig. \ref{fig:dataset}. The 
network thus formed is weighted and undirected.} For convenience, miRNAs are given arbitrary,
but unique, identification numbers $m = 1, 2, \dots i, \dots M,$ where $M$ represents the total number of miRNAs present in the species.  The miRNA network 
is represented as adjacency matrix $W,$ where a element $W_{ij}$ represents the number of mRNAs 
co-targeted by miRNA $i$ and $j$ together. Thus $W_{ij}$ represents the weight of the link joining 
the nodes $i$ and $j.$ If a miRNA pair $i$ and $j$ have no  common targets, they are not connected and
we set $W_{ij}=0.$ The diagonal elements of matrix $W$ are taken to be zero $i. e.,$ $W_{ii}=0.$
The link-weights of the miRNA co-target network can not be ignored while
finding the communities present in the network; the community structure depends on both
weights and the connectivity of the miRNAs. 

\vspace*{.5cm}
{\centering
\begin{minipage}{0.6\textwidth}
\includegraphics[width=8.4cm,bb=0 50 724 570]{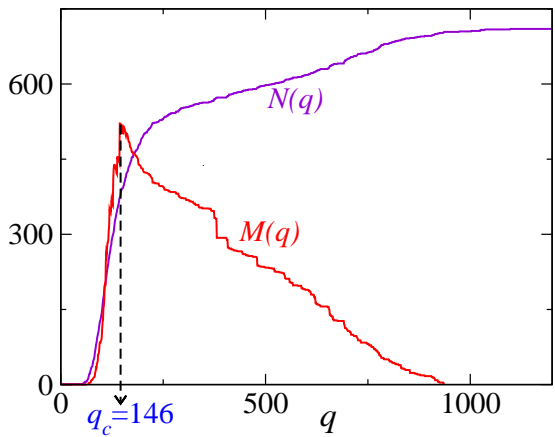}
   \captionof{figure}{The plot of $q$ verses the number of components ($N(q)$) show 
a monotonically increasing curve. At every $q$ the partition of the network correspond 
to a modularity value $M(q)$. The plot of modularity $M(q)$ verses $q$ shows a peak at 
$q=q_c (=146).$}
\label{fig:hsa_mod} 
\end{minipage}
\hspace*{1cm}
\begin{minipage}{0.32\textwidth}
\centering
{ \small
\begin{tabular}{c|c}
\hline
 \textbf{Size} & \textbf{Freq}\\
\hline
 1 & 284\\
 2 & 47\\
 3 & 24\\
 4 & 8\\
 5 & 6\\
 6 & 5\\
 9 & 1\\
 12 & 1\\
 16 & 1\\
 47 & 1\\
 85 & 1\\
\hline
\end{tabular}
}
\captionof{table}{The distribution of size of the components at $q_c$$=$$146$ for 
miRNA co-target network of \textit{Homo sapiens}.}
\label{table:freq_d_1}
\end{minipage}
}

\subsection{Results}
\begin{figure}[h]
\centering
 \includegraphics[width=12cm,bb=0 50 1077 822]{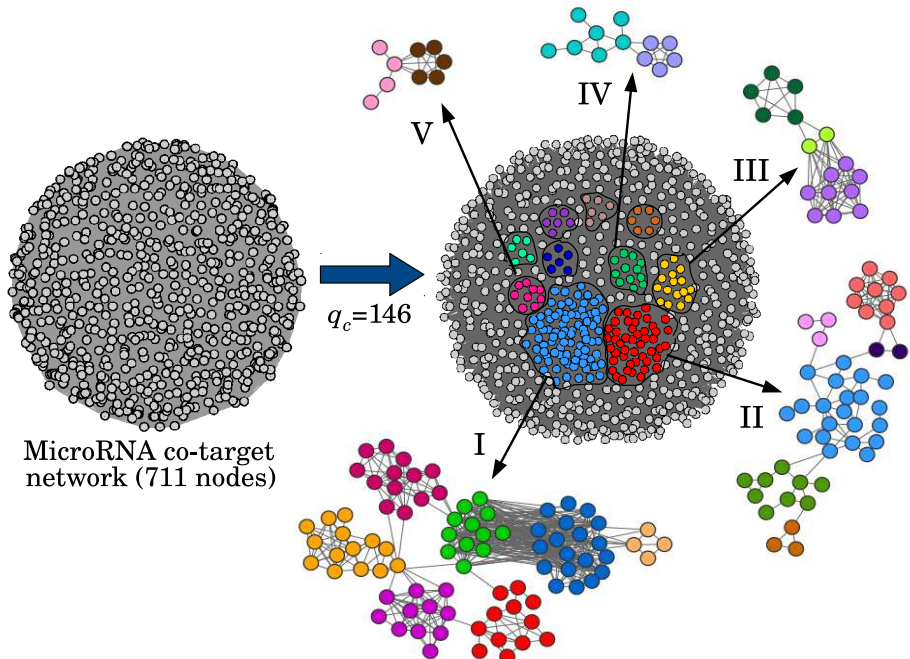}
 \caption{Left:The miRNA co-target network of \textit{Homo sapiens}; it is fully connected network within
$711$ nodes. Right: At $q_c=146$ all the components of size more than $5$ obtained are marked with different colours. Top five components are identified with roman numbers, Component I (size : $85$), II $(47)$, III $(16)$, IV $(12)$ and V $(9)$. These components when further analyzed with NGM weighted algorithm they partition into several modules.}
\label{fig:finestructure}
\end{figure}

We obtain the components of miRNA co-target network by progressively deleting the 
links which have weight less than $q.$ For each $q,$ taking the components as the 
communities of the graph, we calculate modularity $M(q).$ 
Figure \ref{fig:hsa_mod}
shows $N(q)$ and $M(q)$ as a function of $q.$ As expected $N(q)$ is non-decreasing function whereas
 $M(q)$ shows a maximum at $q_c=146.$ Here, the maximum modularity is $M(q_c)=0.044$ and there are
 $379$ components, of which $284$ are isolated miRNAs and  the 
rest $95$ have  two or more miRNAs each (for details refer to Table. \ref{table:freq_d_1}).  
Clearly, most of  the  components   contain
small number  of miRNAs (less than $7$), some  have moderate number $(9,12,16)$
and  only two   are large containing $47$ and $85$  miRNAs.  
 \begin{table}[h]
\centering
 \begin{tabular}{rccccccccccccc}
\hline
\hline
 Module size : & 2 & 3 & 4 & 5 & 6 & 7 & 9 & 11 & 12 & 13 & 14 & 19 & 21\\
  Frequency : & 65 & 38 & 4 & 5 & 1 & 1 & 3 & 1 & 2 & 1 & 1 & 1 & 1 \\
\hline
 \end{tabular}
\caption{Size distribution of miRNA modules obtained using the algorithm proposed in this work. {\clr Note that there are $284$ number of lone miRNAs which are not shown here.} }
\label{table:module_dist}
\end{table}
In the next step we   aim at finding    modules  of all these $95$   disjoint 
 graphs  individually  using NGM algorithm for weighted network \cite{N_wt} to each of them. 
 It turns out that   only  the large  and moderate sized components 
 give rise  to smaller substructures (modules).
For example, the largest  component  (I in Fig. \ref{fig:finestructure}) containing  $85$ miRNAs,  
partitions into $7$ small modules of size $(14, 12, 12, 19, 11, 13, 4)$  and the 
next largest  having $47$ miRNAs (II in Fig. \ref{fig:finestructure})   has  $6$   modules   
of size $(9, 21, 3, 3, 2, 9).$    Partition of other three   components   of size $16, 12$ and $9$
are  also  shown   in Fig. \ref{fig:finestructure} ( marked as III, IV and V respectively).
As a   whole  this algorithm   results in $125$ modules in total. The  distribution of their 
sizes is given in Table \ref{table:module_dist}.

The size of the  partitions obtained for human miRNA co-target network using 
(i) NGM algorithm for weighted network \cite{N_wt}, (ii) clustering algorithm of 
Ref. \cite{susmita} and (iii) the current work are compared in Table \ref{table:compare}. 
It is evident that NGM algorithm gives the highest modularity, but the modules 
obtained there are very large. On the other hand, the clustering algorithm 
\cite{susmita}  gives smaller modularity value and moderate size clusters and 
it was claimed that these clusters are biologically relevant $i. e.,$ they are
pathway, tissue or disease specific. However, some of the clusters are still 
very large, and it is difficult to ascertain functional specificity to these 
clusters. This problem is resolved in our algorithm in expense of low modularity
value. Such partitions can be accepted only when the functional specifications 
obtained here are consistent with those obtained earlier \cite{susmita}.

\begin{figure}[h]
\centering
\includegraphics[width=12.5cm,bb=0 0 1020 454]{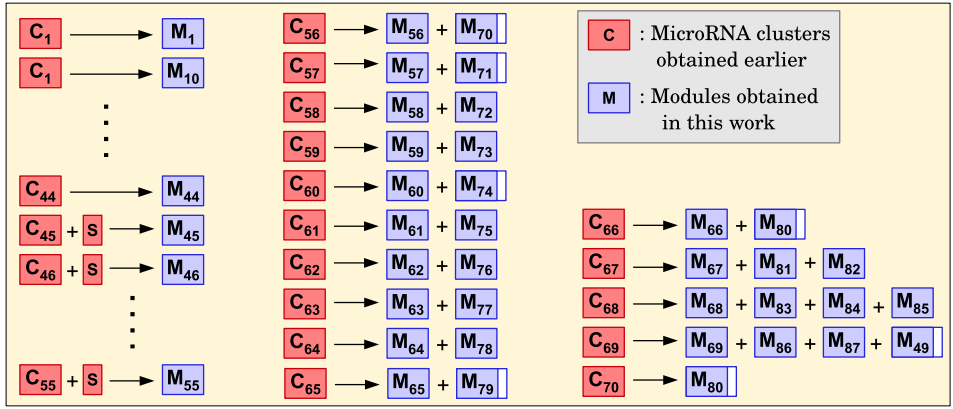}
\caption{Comparison of the modules obtained using the algorithm of this current work with the 
clusters got from the clustering algorithm in Ref. \cite{susmita}. It is clear that all the 
cluster of miRNAs (denoted as $C$) are just combination of the modules (denoted as $M$) 
obtained here. The number written as subscript of $C$ and $M$ represents the 
ID number of the clusters and modules.}
\label{fig:comp_module}
\end{figure}

{\clr In Ref. \cite{susmita} the authors have obtained $70$ clusters, each having 
two or more miRNAs. All these  clusters  are found to be  pathways, disease  or 
tissue specific; for convenience, we denote them as $C_1, C_2, \dots ,C_{70}.$  
We analyze the miRNA contents of these $70$ clusters in terms of the $124$ modules 
obtained in this work (namely $M_1,M_2, \dots M_{124}$).} If modular structure 
of miRNAs are  different from those of the clusters, one would expect that each 
cluster would contain miRNAs belonging from many different modules. However we 
find that each cluster, in terms of their miRNA content, is  either identical to 
one of the modules or composed of at most four modules. This is described in  
Fig. \ref{fig:comp_module} in details. As described in the Fig. \ref{fig:comp_module}, clusters 
$C_1$ to $C_{44}$ are identical to the respective modules $M_1$ to $M_{44}$. 
Module $M_{45}$ is same as $C_{45}$ but contains one extra miRNA, marked as $S$
in Fig. \ref{fig:comp_module}; the same is true for modules $M_{46}$ to $M_{55}.$
MicroRNAs of all other clusters $C_{56}$ to $C_{70}$ comes from two or more modules.  
If all miRNAs of a module participate in forming a cluster we represent it in 
Fig. \ref{fig:comp_module} by a fully shaded box, or otherwise by a partially 
shaded box. For example, $C_{60}$ consists of all miRNAs of module $M_{60}$  and  
some miRNAs of $M_{74}.$ Note that miRNAs of module  $M_{49}$ belong to two 
clusters $C_{49}$ and $C_{69}$;  another example is $M_{80},$ whose miRNAs belong 
to $C_{66}$ and $C_{70}.$ 
\begin{savenotes}
 \begin{table}
 \centering
 \begin{tabular}{|l|llc|}
 \hline
 Methods  & $N(q)$ & {\clr Component size} & $M(q)$ \\
 \hline
 NGM algo. \cite{N_wt}  & $4$ & $(6, 79, 294, 332)$ & $0.081$ \\
 Clustering algo. \cite{susmita} & $94$ \footnote{MiRNAs of the giant cluster ${\cal G}$ in 
Ref. \cite{susmita} consists of $70$ modules; the rest of the miRNAs form $24$ modules.} & $(1, 2, 3, 4, 5, 6, 7, 8, 9, 11, 14, 16, 31, 47)$ & $0.025$ \\
This work & $124$ & $(1, 2, 3, 4, 5, 6, 7, 9, 11, 12, 13, 14, 19, 21)$ & $0.022$ \\
\hline
\end{tabular}
\caption{Comparison of the three methods {\clr in context of} finding the modules of densely 
connected weighted
miRNA co-target network. The number of components or modules $N(q)$ obtained with the corresponding
modularity $M(q)$ are mentioned along with the sizes of the components for each of the algorithms.}
\label{table:compare}
\end{table}
\end{savenotes}
This analysis revels that the modules obtained in  this  work  are either same or very 
similar to those obtained in \cite{susmita}.    Since 
miRNA  clusters     are known to be  pathway and tissue specific, 
the modules  obtained here,   which are combined  to form  the clusters are also
biologically relevant \cite{aaa}.

\section{Conclusion}
In this article we propose an algorithm to detect community structure of dense 
weighted networks. If the network has adjacency matrix $W$ whose elements $W_{ij}$ 
refer to the weight of the link connecting nodes $i$ and $j$, one can implement 
the algorithm by the following steps, 
{\bf I.} Delete  all the links having weight $W_{ij}<q$; find the modularity $M(q)$
 of the network taking the disjoint components obtained here as the partitions. 
{\bf II.} Find $q_c$ where $M(q)$ is maximum. 
{\bf III.} Take all the components at $q=q_c$ containing two or more miRNAs, one at a 
time,   apply Newman Girvan's weighted algorithm to obtain its modules. 
To demonstrate the algorithm, we consider miRNA co-target network of \textit{Homo sapiens},
which is dense and weighted, and compare the modules with the miRNA clusters 
obtained earlier \cite{susmita}. It turns out that most clusters are either 
identical to one of the  modules, or composed of miRNAs belonging to at most 
four different modules. Thus, like the clusters, modules are also involved in
specific biological functions. 
 
This algorithm has certain advantage over some of the standard ones. The NGM algorithm for 
weighted networks \cite{N_wt} can not resolve small sub-structures if the network is dense.
The algorithm of Ref. \cite{susmita} can overcome this difficulty, but does not resolve 
communities which are interlinked by a few links  having very large weights. The algorithm 
discussed here combines both the methods suitably and overcome their shortcomings. Unlike 
the algorithm of \cite{susmita}, where actual number of clusters depends (though weakly) 
on the final choice of $q$($=160$ in \cite{susmita}) this algorithm is free from parameters
and provide an unique partition of a weighted network. 
 
It has been known that a network containing $l$ connections can not resolve any module
which has $\sqrt{l/2}$ links. Usually, a densely connected weighted network, with a wide 
distribution of link weights falls in this category and it is difficult to resolve small
substructures of these networks. We believe the algorithm considered here is general, 
though discussed in context of miRNA co-target networks, and can be used for community 
detection in dense and weighted networks. 

\section*{Acknowledgements}
The author would like to gratefully acknowledge P. K. Mohanty for his constant encouragement and 
careful reading of the manuscript. His insightful and constructive comments have helped 
us a lot in improving this work.


\begin{thebibliography}{99}
\bibitem{aaa} An independent study of biological relevance of these  modules will be reported elsewhere.
\end{thebibliography}
\end{document}